\documentclass[reprint,prb,aps,twocolumn,showpacs,superscriptaddress]{revtex4-1}
\usepackage{amsmath}
\usepackage{graphicx}
\usepackage{dcolumn} 
\usepackage{bm}
\usepackage{epsfig}
\usepackage{xcolor}

\begin{document} 
\title{\textit{Ab initio} study of nontrivial topological phases in corundum structured $($M$_2$O$_3)/($Al$_2$O$_3)_5$ multilayers.}

\author{Juan F. Afonso}
\email{jf.afonso@usc.es}
\affiliation{
Departamento de F\'{\i}sica Aplicada, Universidade de Santiago de Compostela, E-15782 Santiago de Compostela, Spain
}
\affiliation{
Instituto de Investigaci\'{o}ns Tecnol\'{o}xicas, Universidade de Santiago de Compostela, E-15782 Santiago de Compostela, Spain
}
\author{Victor Pardo}
\email{victor.pardo@usc.es}
\affiliation{
Departamento de F\'{\i}sica Aplicada, Universidade de Santiago de Compostela, E-15782 Santiago de Compostela, Spain
}
\affiliation{
Instituto de Investigaci\'{o}ns Tecnol\'{o}xicas, Universidade de Santiago de Compostela, E-15782 Santiago de Compostela, Spain
}


\begin{abstract}

\textit{Ab initio} calculations have been performed on hexagonal layers of M$_2$O$_3$ (M being several transition metals of the $5d$ series) sandwiched by a band insulator such as Al$_2$O$_3$ that provides the honeycomb lattice where the $5d$ electrons reside. This corundum-structure-based superlattice is the most obvious way to design a honeycomb lattice with transition metal cations avoiding the use of largely polar surfaces. We obtain that this system supports the presence of Dirac cones at the Fermi level that open up with the introduction of spin-orbit coupling at various fillings of the $5d$ band.  The DFT calculations performed in this work show that the $5d^5$ situation is always a trivial insulator, whereas the $5d^8$ filling presents topological insulating configurations which evolve into a trivial state with increasing tensile strain or on-site Coulomb potential U. However, LDA+U calculations show a stable antiferromagnetic solution for the $5d^8$ case at every U value, which would break time reversal symmetry and could affect the topological properties of the system. We also discuss the similarities with the buckled honeycomb lattice obtained using perovskite (111) bilayers previously studied in literature, in particular for the $5d^5$ and $5d^8$ configurations. This work provides some clues on the stability of topological phases using metal oxides in general.

\end{abstract}

\maketitle



\section{Introduction}\label{intro}

In recent times, the interest in topological insulators\cite{TI_moore} has become ubiquitous in condensed matter physics literature.\cite{reviewti, QSH_TI_Zhang, ti_3d_fu_kane_mele} The first observed examples were found in the classic thermoelectric materials\cite{TI_termoelec} and their nanostructures,\cite{qw_hg_te_th, qw_hg_te_exp} but soon the list expanded to include oxides,\cite{ti_oxide_perovskite, okamoto_111_perovskite, xiao_111_perovskite, xiao2, joselado_perovskites} compounds based on Bi, Te and Sb,\cite{bi_1_x_sb_x, bi_te_sb_dirac_cone, bi2te3_ti_exp_3d, ti_la_bi_te_ab_initio, tlbise2_dirac_exp, bi2te2_res_quant_transport, bi_se_te_compounds_exp} and even Kondo-type insulators.\cite{kondo_ti, kondo_ti_2, kondo_cubic_ti, kondo_smb6_th, kondo_smb6_exp} Various types of unique topological phases have been explored, including the interface between a topological insulator and a superconductor,\cite{junction_ti_supercond, junction_supercond_ti_supercond, ti_supercond_qbts} topological ordering in superconductors,\cite{ti_supercond_observation} the existence of so-called topological Mott insulators,\cite{ir-top-mot-ins, top-mot-ins, top-mot-ins-1D} topological crystalline insulators,\cite{tci_th, tic_th_2, tic_pb_1-xsn_xse} and the list continues. We are on the way to start thinking about building devices such as field effect transistors out of topological insulators,\cite{ti_FET, ti_trans} and yet their properties are still intriguing and good candidates for applications in electronics (\textit{e.g.} having a large band gap\cite{Felser_BaBiO3}) or in quantum computing (\textit{e.g.} by producing Majorana fermions\cite{prespective_majorana, makorana_fu_kane}) still are to be found experimentally. Topological insulators (TI) constitute a new state of matter that has forced us to rewrite our basic solid state physics textbooks. They are insulating materials in their bulk state but at the same time they present dissipationless chiral channels on the surface. Those edge states present an absence of back scattering that opens new possibilities for charge and spin transport in nanostructures. They are also topologically protected by certain symmetries of the one-particle Hamiltonian that describes the properties of the system. The quest in under way to find new topological insulators that could be stable and usable. One possible path is to try to design them based on oxides. These have the advantage of being very well known materials, stable in ambient conditions and can be grown in various nanostructures, such as thin films or multilayers using standard techniques. One such route has been explored recently by various authors.\cite{xiao_111_perovskite, joselado_perovskites, Pentcheva_LaNiO3, piro_1, piro_2} In particular, creating a honeycomb lattice using a perovskite bilayer\cite{xiao_111_perovskite} has produced interesting predictions for various $5d$ transition metal compounds being the active part of the multilayered system. We propose here an analogous design but building the honeycomb lattice in a more natural way, avoiding the use of a largely unstable surface of the perovskite, the highly polar (111) surface that only recently has started to become accessible for experimentalists.\cite{111_perovskite_exp, 111_srfeo3, 111_lanio3, review_epitaxial,eondu_cb} We will use instead a non-polar surface based on the corundum structure\cite{corundum_pauling_1925} that provides intrisically a transition metal-based honeycomb lattice.

\begin{figure*}[htc!]
\includegraphics[height=7cm]{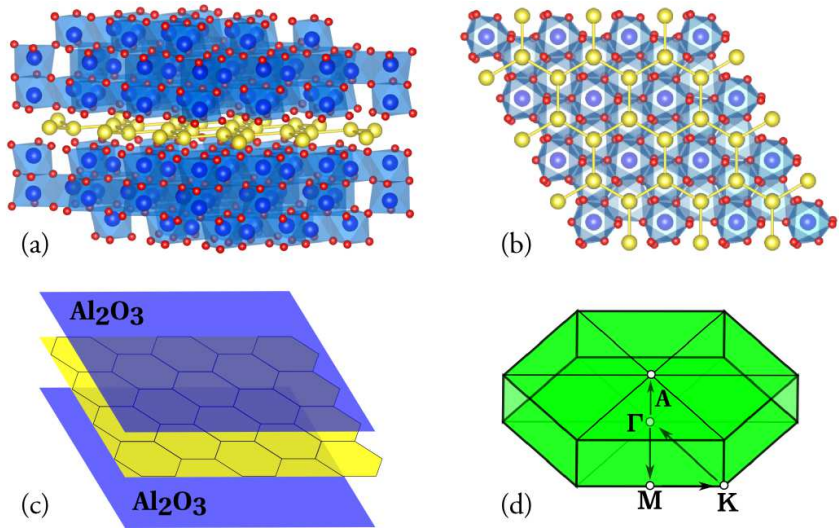}
\caption{(Color online) (a) and (c) Scheme of the corundum structure multilayer (M$_2$O$_3$)/(Al$_2$O$_3$)$_5$ studied where the active single layer M$_2$O$_3$ is highlighted. (b) Top view of the honeycomb lattice formed by the $5d$ TM cations $M$. (d) Reciprocal lattice showing the path of the band structures presented.}
\label{corundum_structure}
\end{figure*}

The paper is organized as follows: in Section \ref{comp} we will present the details of the \textit{ab initio} calculations, in Section \ref{struct} we will describe the system under study, in Section \ref{results} we will analyze the results of our calculations for the most interesting compounds, suggesting the analogies and differences with results already available in literature: Section \ref{results_gga} describes the $5d^5$ and $5d^8$ systems as possible candidate band structures to host non-trivial topological properties, in Section \ref{results_correlations} we provide a description of the magnetic properties of these systems, in Section \ref{detailed} we provide a detailed study of each case as a function of strain and in Section \ref{results_lda_U} we include the effects of correlations at the LDA+U level. Finally, in Section \ref{conclusion} we will summarise the main implications of our work.


\section{Computational details}\label{comp}

\textit{Ab initio} electronic structure calculations based on the density functional theory\cite{hk,ks} (DFT) have been performed using the all-electron full potential {\sc WIEN2k}\cite{wien} software. 
The calculations were performed for different off-plane and in-place lattice parameters of the corundum structure, all of them were relaxed using the full symmetry of the original cell. The exchange-correlation term depends on the case. Uncorrelated calculations were performed using the Generalized Gradient Approximation (GGA) with the Perdew-Burke-Ernzerhof scheme.\cite{gga} The Local Density Approximation+U (LDA+U) in the so-called ``fully localized limit"\cite{sic1} was utilised in order to explore the effects of on-site Coulomb repulsion in the $5d$ manifold. For selected configurations, GGA\cite{gga} and LDA\cite{lda} calculations are shown to yield qualitatively equivalent results. This will be presented and discussed in Section \ref{results}.

The calculations were performed with a k-mesh of $8\times8\times8$ and R$_{mt}$K$_{max}= 7.0$ (the product of the smallest muffin-tin radius and the maximum k-vector of the plane-wave expansion determines the size of our basis set). Spin-orbit coupling (SOC) was introduced in a second variational manner by using the scalar relativistic approximation,\cite{singh} these calculations were performed with a modified k-mesh of $8\times8\times2$. In the GGA calculations of the bulk structures of M$_{2}$O$_{3}$ of Os and Au, the muffin-tin radii R$_{mt}$ (in \textit{a.u.}) were set to $1.92$ for the metal M and $1.57$ for the O. In the $(\text{Au}_2\text{O}_3)/(\text{Al}_2\text{O}_3)_5$ multilayer, the R$_{mt}$ values used were: 1.78 for Au, 1.68 for Al and 1.5 for O;  for the $(\text{Os}_2\text{O}_3)/(\text{Al}_2\text{O}_3)_5$ multilayer: 1.78 for Os, 1.68 for Al and 1.58 for O; in the $(\text{Ir}_2\text{O}_3)/(\text{Al}_2\text{O}_3)_5$ multilayer: 1.92 for Ir, 1.61 for Al and 1.61 for O.


\section{The system}\label{struct}

We can see in Fig. \ref{corundum_structure} the multilayered structure of the system under study, where a single layer of M$_2$O$_3$, M being a $5d$ transition metal (TM), is sandwiched by a wide gap band insulator, such as Al$_2$O$_3$,\cite{lattice_band_structure_al2o3,surface_oxides_book} with the so-called corundum structure that provides the symmetry of the multilayer. In that geometry (Fig. \ref{corundum_structure} (b)), the TM cations sit on a honeycomb lattice, resembling the situation in graphene or in other oxide-based heterostructures, such as the aforementioned perovskite bilayers grown along the (111) direction, but without the additional bucklings caused by the bilayer structure, it is a single-layer here that provides the honeycomb. For that reason, this system is a more natural realization of a $5d$ metal-based honeycomb lattice utilising non-polar oxides for its growth. If a similar electronic structure occurs, then comparisons can be drawn with other similar systems that sustain topologically non-trivial phases. We provide a description of these in the following Section.


\section{Results}\label{results}

\subsection{General overview of the electronic structure}\label{results_gga}

\begin{figure}[hc!]
\includegraphics[width=.8\columnwidth]{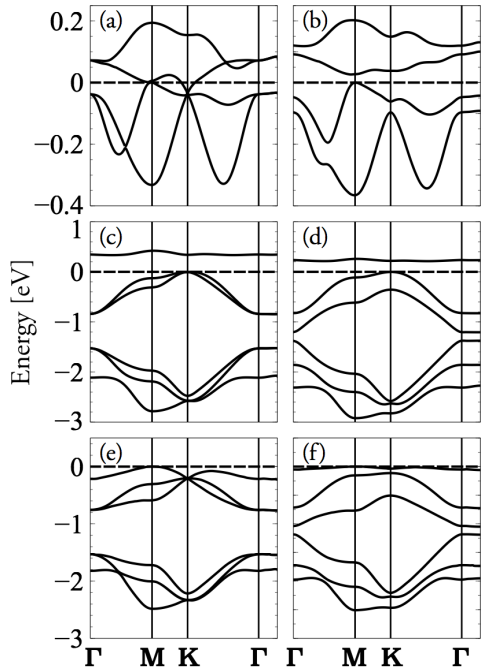}
\caption{(a) and (b) show the band structure obtained within GGA for the case without (left)/with(right panel) spin-orbit coupling, respectively, for the $5d^{8}$ filling (Au$_2$O$_3$) at $a = a_{Al_2O_3}$ and optimized \textit{c} lattice parameter $c = 26.51$ \AA. Only the four $e_g$ bands are shown. (c) and (d) show similar calculations for the $5d^{5}$ filling (Os$_2$O$_3$) with $c = 25.78$ \AA, where only the t$_{2g}$ states are shown. (e) and (f) show the $t_{2g}$ manifold for the $5d^{6}$ filling (Ir$_2$O$_3$) with $c = 25.50$ \AA, where the Dirac cone opens up with SOC at the $5d^5$ filling.}
\label{au_gga_soc_vs_no_soc}
\end{figure}

DFT calculations were performed for various $5d$ TM cations on the $\text{M}_2\text{O}_3$ sandwiched layer. We can see in Fig. \ref{au_gga_soc_vs_no_soc} the band structure of the multilayer around the Fermi level with and without spin-orbit coupling for various fillings of interest: half-filled $e_g$ levels (Au$_2$O$_3$, panels (a) and (b)), one hole in the $t_{2g}$ multiplet (Os$_2$O$_3$, panels (c) and (d)) and $5d^{6}$ configuration (Ir$_{2}$O$_{3}$, panels (e) and (f)) that has a full $t_{2g}$ shell but shows a Dirac cone at $5d^{5}$ filling. We have studied all the other possible fillings but none of them show anything close to a Dirac point near the Fermi level. 

In the perovskite-based bilayer system, two of these fillings ($5d^{8}$ and $5d^{5}$) have been predicted to be topologically non-trivial.\cite{xiao_111_perovskite, joselado_perovskites} Here we see a similar band structure. However, for the $5d^8$ case, a band crossing at the Fermi level at the \textbf{M} point is removed by SOC, whereas in the perovskite bilayer case it occurs at \textbf{K}.\cite{xiao_111_perovskite,joselado_perovskites} This suggests that the tight binding model for this unit cell is different. This is consistent with the different hopping paths in the corundum unit cell in comparison with the perovskite case. Yet, it is worth analysing the similarities in terms of their topological properties.

\begin{figure*}[htc!]
\includegraphics[height = 5.5cm]{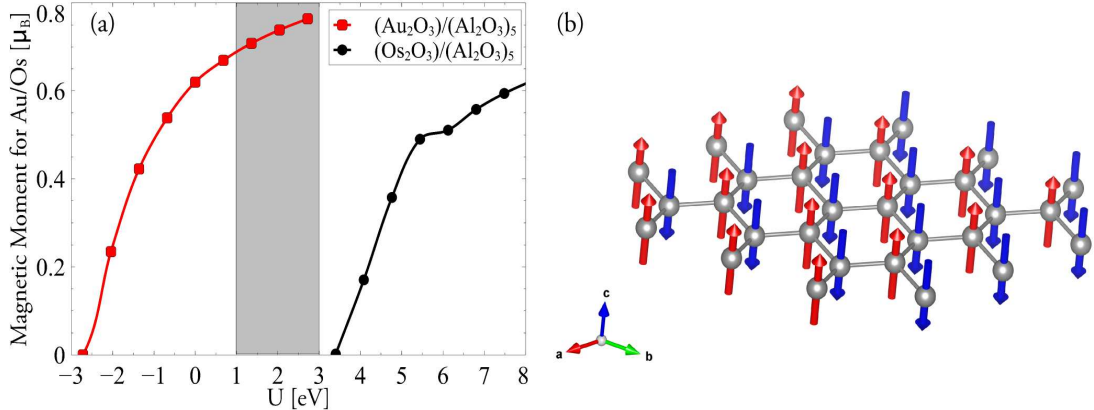}
\caption{(Color online) (a) Representation of the magnetic moment evolution with U, both for $5d^5$ (black circles)  and $5d^8$ (red squares) fillings. The shaded region highlights the magnetic moment calculations inside the typical realistic values of U for $5d$ electrons. Notice the similarity in the evolution of the magnetic moments between the two systems studied, but just at two different U regimes. (b) Schematic representation of the AF solution in the honeycomb plane.}
\label{MMI_U}
\end{figure*}

Also, for the $5d^5$ case, we notice that the bands are considerably broader and hence the relative strength of SOC will be smaller compared to the perovskite-bilayer case. We will explore below what happens when the SOC strength is artificially enhanced. Moreover, for the Os-based multilayer we see the gap is not affected by SOC. However, it looks like the system can be close to a topological transition via increasing SOC, reducing hopping parameter $t$, applying strain, etc. since the Ir-based multilayer shows a Dirac cone at $5d^5$ filling which responds to SOC by opening a gap as shown in Fig. \ref{au_gga_soc_vs_no_soc} (e) and (f). We will explore further all these possibilities throughout the paper.

Due to the presence of time reversal symmetry (TRS) and inversion symmetry (IS), the topological invariants of the band structure can be easily computed in the bulk state and we can determine whether the system is a TI or a trivial insulator by following the simple prescription of Fu and Kane. This computes the Z$_2$ invariant of the system utilizing only the single-particle eigenstates at the time-reversal-symmetric k-points of the Brillouin zone,\cite{fu_kane_method} as long as TRS and IS are retained, the latter occurring for all the calculations presented.





By means of this rule, we will be able to analyse in detail whether this multilayered system (when it is TR symmetric) is topologically trivial or non-trivial, depending on the Z$_2$ invariant that results from the calculations.

\subsection{Importance of magnetism and correlations}\label{results_correlations}

Before we apply the so-called Fu-Kane\cite{fu_kane_method}, which is only valid when TRS and IS are present, let us recall that this requieres that the system must have a non-magnetic solution, since IS is always retained in our calculations. We should clarify in what regime these multilayers will present a non-magnetic solution as most stable. For that sake, we have studied the evolution of magnetism with the on-site Coulomb repulsion U by means of the LDA+U method for the $($M$_2$O$_3)/($Al$_2$O$_3)_5$ system by setting up a nearest neighbor antiferromagnetic (AF) coupling between the transition metals in the honeycomb plane (Fig. \ref{MMI_U} (b)). Being rather delocalized, the 5d electrons in these systems are usually treated with values of the on-site Coulomb repulsion on the order of $1-3$ eV, the correct values being difficult to estimate, we present results calculated at various U values. 

The evolution of magnetism with the Coulomb interaction is summarised in Fig. \ref{MMI_U} for the two fillings we will focus along the paper in. We see that as U increases, an AF solution becomes stable for both compounds, the onset point being at different values of U.

The Os-based system presents an AF ordering for $\text{U} > 3.5$ eV (a sizable value for these 5d-electron systems) that would only enhance the gap without producing any crossover to a phase with different topological properties. In contrast, the Au-based system has a stable AF solution for any $\text{U} \geq 0$. Even though our DFT calculations predict an AF phase to be stable, let us point out that Au$_2$O$_3$ in the bulk form is itself non-magnetic.\cite{au2o3_magn} The same non-magnetic result is obtained from our own calculations for bulk Au$_2$O$_3$ with corundum structure. Also, another Au$^{3+}$ compound like LaAuO$_3$ was found experimentally to be diamagnetic.\cite{laauo3_diam} Thus, it is interesting here to study the topological properties of a non-magnetic phase as well, as we will carry out in detail below. Yet, more studies are required to understand if a hexagonal single-layer Au$_2$O$_3$ would sustain magnetism or not. In case it does, such system would be very interesting to study the effects of strong correlations in those very narrow bands produced by spatial confinement. Previous works based on perovskites bilayers\cite{joselado_perovskites, okamoto_correlation_effects} have also studied the magnetic solutions of the $5d^5$ and $5d^8$ compounds. For the $5d^5$ filling case,\cite{joselado_perovskites} an AF solution becomes stable for $\text{U}>3$ eV breaking TR and IS which is similar to the results obtained for the corundum-based system presented here. On the other hand, the $5d^8$ system\cite{okamoto_correlation_effects} presents a narrow window inside the AF region ($\text{U} > 2.1$ eV) where the system remains in a topological insulating state. By increasing the parameter U, this phase is destroyed and the system evolves (through gap closing) into a trivial insulator with AF ordering. In this corundum-based system, the e$_g$ bandwidths are substantially reduced with respect to the perovskite-based case, promoting the strength of the magnetic coupling. Yet, in principle magnetism need not break the topological insulating phase. Ref. \onlinecite{joselado_perovskites} shows that breaking inversion symmetry does not immediately destroy the topological insulating phase in the case of perovskite bilayers either. The actual topological features of the AF phase for these corundum-based systems are beyond the scope of this paper.


\subsection{Detailed analysis case by case.}\label{detailed}

Once we have clarified what DFT calculations predict about the stability of magnetism in these systems, we focus on analyzing the topological properties of non-magnetic phases and their strain dependence carrying out calculations at different values of U but retaining a non-magnetic solution where TRS is not broken.

\begin{figure*}[htc!]
\includegraphics[height=8.5cm]{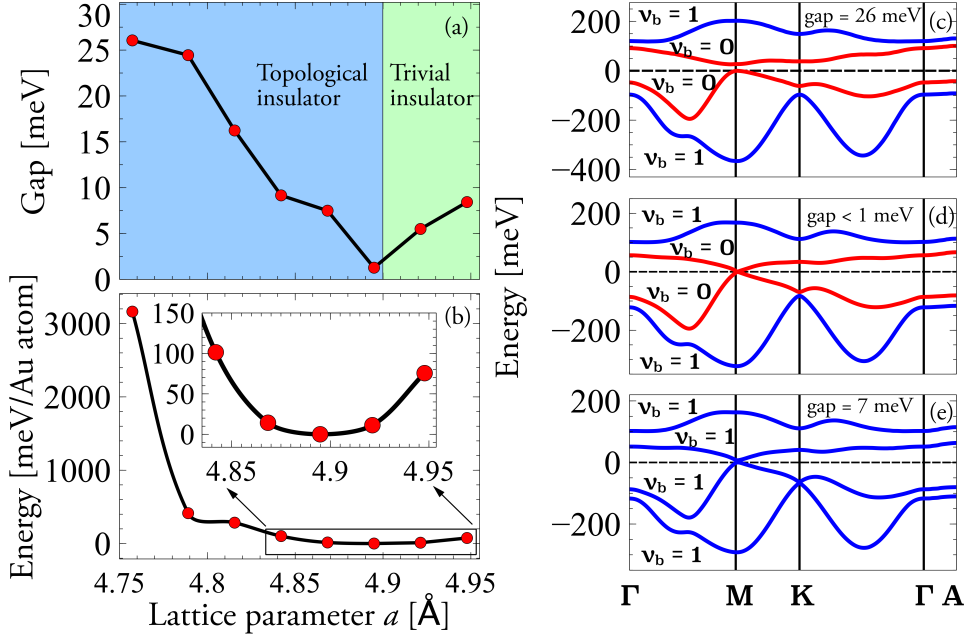}
\caption{(Color online) Evolution of the band gap (a) and the total energy (b) for the $(\text{Au}_2\text{O}_3)/(\text{Al}_2\text{O}_3)_5$ system.  All the GGA calculations were performed at the optimized lattice parameter \textit{c} with relaxed atomic positions. The inset zooms at the region close to the energy minimum. Band structure evolution (only the $e_g$ manifold is shown) for the $5d^{8}$ filling with strain: (c) $a_{Al_2O_3} = 4.76$ \AA, (d) $a_{GS} = 4.89$ \AA\ and (e) $a = 4.95$ \AA\ as the systems evolves from a topological to a trivial insulator by closing and reopening the gap. The topological invariants of each band are calculated and shown.}
\label{gga_au}
\end{figure*}

\subsubsection{$(\text{Au}_2\text{O}_3)/(\text{Al}_2\text{O}_3)_5$ system}

\begin{figure*}[htc!]
\includegraphics[height=5cm]{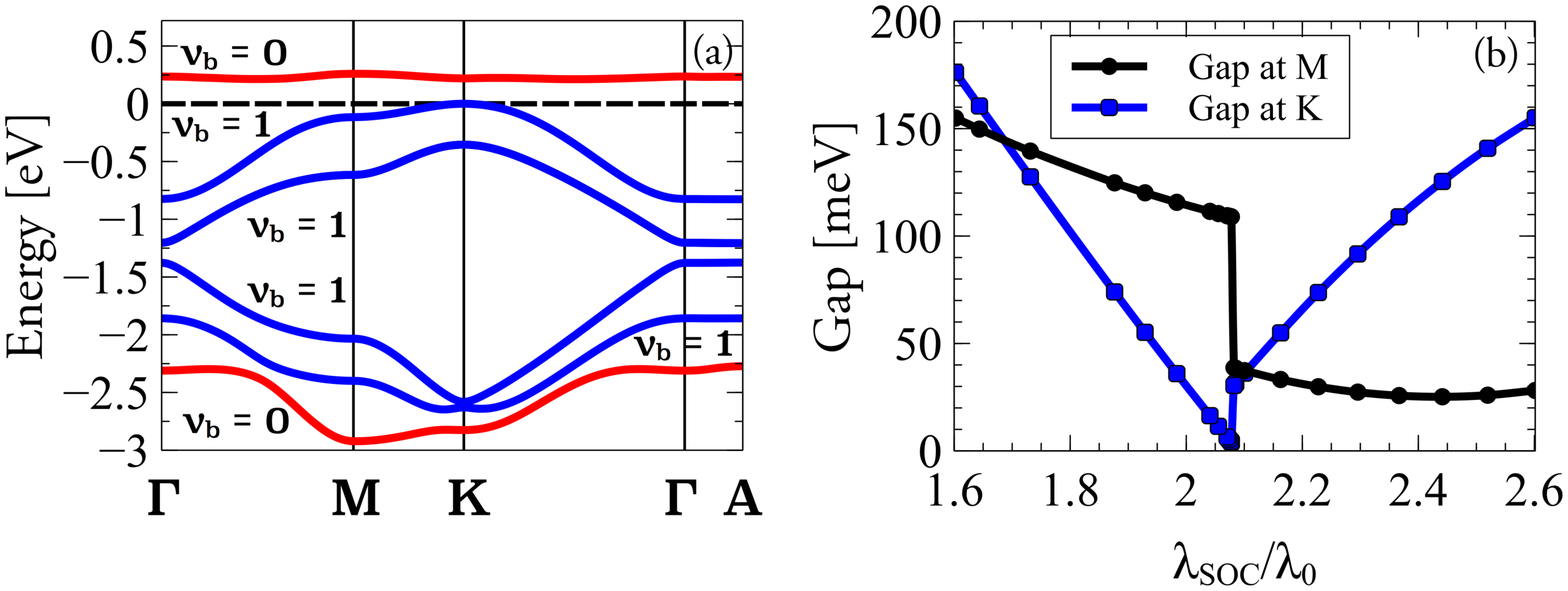}
\caption{(Color online) (a) Band structure calculated with spin-orbit coupling showing the $t_{2g}$ manifold of the system $(\text{Os}_2\text{O}_3)/(\text{Al}_2\text{O}_3)_5$ for the ground state configuration with $a = 4.80$ \AA\ and $c = 13.64$ \AA. The topological invariants calculated show a trivial insulating state that holds for all the lattice parameters studied. Observe that the gap is sizable (about 200 meV) in this case, too big to be easily reduced by applying strain. (b) Evolution of gap with the SOC strength parameter $\lambda_{so}/\lambda_{0}$ (the artificially modified value divided by the actual one) at k-points \textbf{M} and \textbf{K}. The gap becomes zero at approximately $\lambda_{so}/\lambda_{0} \approx 2.1$ where the gap presents a discontinuity at both k-points studied. Several calculations with slightly different values of $\lambda_{so}/\lambda_{0}$ around $2.1$ were performed but no intermediate value of the gap in the interval was observed, there is a discontinuous change there at both k-points simultaneously. Even though there is a drastic phase change observed, all the calculations preformed show a trivial insulating phase for $(\text{Os}_2\text{O}_3)/(\text{Al}_2\text{O}_3)_5$.}
\label{gga_os}
\end{figure*}

Let us first discuss the properties for the $5d^8$ filling. This is a half-filled $e_g$ shell, with all the $t_{2g}$ levels completely occupied and far below the Fermi level. We will forget about them in our description.

 Figure \ref{gga_au} (a) shows the evolution of the gap and the total energy of the system (Fig. \ref{gga_au} (b)) with in-plane strain (lattice parameter $a$). The picture shows how, as $a$ increases, the topological gap decreases to zero and reopens again as a trivial insulator. The total energy curve shows that the crossing point (gap zero) would be very close to the equilibrium structure. Due to the lattice size mismatch between Al$_2$O$_3$ and Au$_2$O$_3$ (which is quite substantial; a GGA-based structural optimization in bulk Au$_2$O$_3$ with corundum structure yields $a = 5.48$ \AA\ and $c = 15.6$ \AA, that is a sizable in-plane mismatch of about 15 \%), the lattice prefers to deform its volume increasing the $a$ lattice parameter compared to that of Al$_2$O$_3$. If these multilayers can be grown epitaxially retaining the in-plane lattice parameter of the substrate (Al$_2$O$_3$, $a = 4.75$ \AA), the system would be well inside the topologically non-trivial region of the phase diagram. However, our total energy calculations suggest a more stable situation would occur closer to the transition point with a much smaller band gap, and also that epitaxial growth on a substrate with larger \textit{a} could tend to reduce the topological gap. In any case, tensile strain drives the system towards a trivial insulating phase whereas compressive strain induces a topological insulating phase.

Figure \ref{gga_au} (c)-(e) shows the evolution of the band structure with strain. We see there the band structure for three different values of the in-plane $a$ lattice parameter, panel (c) with the $a$ of Al$_2$O$_3$ and panels (d) and (e) for two larger $a$ values. For each $a$ parameter, the $c$-lattice parameter was optimised. We see that the gap at the \textbf{M} point closes as $a$ increases and then reopens again. When it does so, the $\nu$ value of the bands around the Fermi level has changed. By computing the Z$_2$ invariant following the Fu-Kane criterion, we see that the system is a topological insulator for $a= 4.75$ \AA, the gap becomes zero at around $a = 4.89$ \AA, and it reopens as a trivial insulator at larger $a$ values. 

\subsubsection{$(\text{Os}_2\text{O}_3)/(\text{Al}_2\text{O}_3)_5$ system}

A similar analysis has been carried out for the $(\text{Os}_2\text{O}_3)/(\text{Al}_2\text{O}_3)_5$ case. In this $5d^5$ filling (one hole in the $t_{2g}$ manifold), corresponding to the nominal occupation of the Os$^{3+}$ cation, the existence of a topological insulating phase is not observed. All the configurations studied were found to be trivial insulators (or metallic phases). We have optimized the atomic positions and the volume ($a$ and $c$ lattice parameters) and the ground state was found for $a = 4.80$ \AA, very close to the value for Al$_2$O$_3$. A wide range of both tensile and compressive strains were tested by varying the in-plane $a$ lattice parameter (optimizing $c$ in every case). None of the strains considered yielded a topological insulating phase. However, we observed that a Dirac cone at the \textbf{K} point near the Fermi level that opens up with SOC exists for the $5d^5$ filling when the active layer is Ir$_2$O$_3$, nominally a $5d^6$ state (if the chemical potential were shifted to one electron less). This could indicate that the system is close to a transition towards a topologically non-trivial solution, which we have tried to explore. This $5d^5$ filling has been predicted to host topologically non-trivial phases for the buckled honeycomb lattice constructed using a perovskite bilayer along the (111) direction. We can see that the situation changes here for Os$_2$O$_3$ as an active layer, whose band structure including spin-orbit coupling can be seen in Fig. \ref{gga_os} (a). This shows only the $t_{2g}$ states, with one unoccupied band just above the Fermi level. The band width of the $t_{2g}$ multiplet is quite large, of about 3 eV, substantially larger than in the perovskite bilayer system that hosts topologically non-trivial phases (where the band width is about 1.5 eV).\cite{xiao_111_perovskite, joselado_perovskites} In the corundum case, the hoppings occur all of them in one plane (compared to the buckled honeycomb lattice in the case of the perovskite bilayers) and broader bands show up. In both cases, the top of the valence band is at the \textbf{K} point. The fact that the ratio $\lambda_{\text{so}}/t$ (being $\lambda_{\text{so}}$ the SOC strength parameter) is small implies that SOC is not strong enough to promote a band inversion that could lead to a topologically non trivial phase. Still, two-dimensional confinement leads to an insulating phase which survives for a significant range of strain effects.

In order to test how far this system is from a spin-orbit induced phase transition we will enhance the strength of the SOC and calculate the electronic structure for various values of $\lambda_{\text{so}}$, where the standard spin-orbit term in the Hamiltonian was considered in a scalar relativistic approximation.

\begin{figure*}[htc!]
\includegraphics[height=6.5cm]{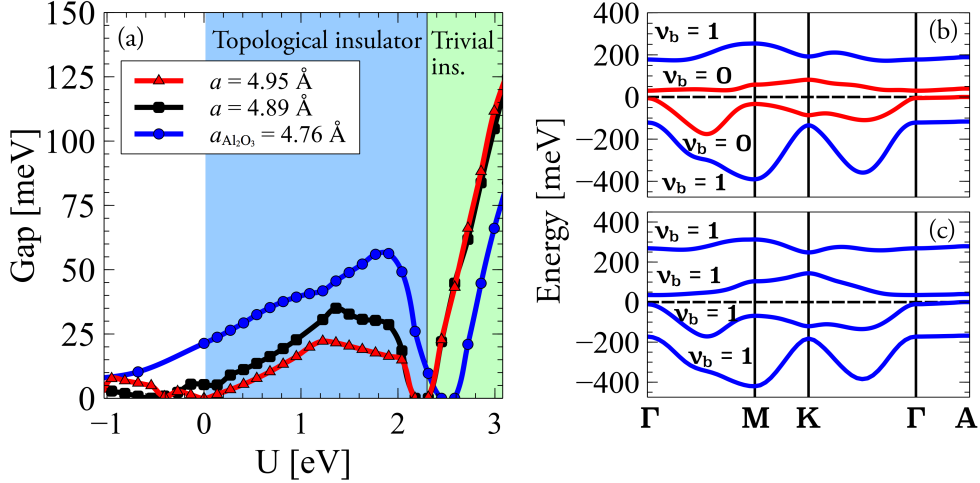}
\caption{(Color online) (a) Evolution of the gap with the on-site Coulomb repulsion U for the system $(\text{Au}_2\text{O}_3)$ with $5d^8$ filling for the lattice parameters indicated. A transition between a topological insulator an a trivial one is produced at $\text{U}=2-3$ eV for the studied configurations. (b) Band structure calculation for the Au$_2$O$_3$-based multilayer ($5d^{8}$ filling) with $a= 4.76$ \AA\ for $\text{U}=2.31$ and $\text{U}=2.72$ eV. The increase in U induces a change in the topological properties of the system from a low-U topological phase to a higher-U trivial insulating solution. All of the LDA+U calculations performed show a non-magnetic solution.}
\label{lda_u_au}
\end{figure*}

The enhancement of the SOC strength was performed for the ground state configuration of this system with lattice parameters $a = 4.80$ \AA\ and optimized $c$. The evolution of the gap at both k-points \textbf{M} and \textbf{K} with the ratio $\lambda_{\text{so}}/\lambda_0$ ($\lambda_0$ being the physical value of the SOC strength) is presented in Fig. \ref{gga_os} (b). The gap becomes smaller with increasing SOC strength and it closes and reopens again at $\lambda_{\text{so}}/\lambda_0  \approx 2.1$. The topological invariants of the band structure $\nu_b$ remain the same at both sides of the zero-gap point. As a consequence of this, the insulating phase remains trivial and no TI state was found. However, a discontinuous change in the gap occurs at that $\lambda_{\text{so}}/\lambda_{0}$, as we can see in Fig. \ref{gga_os} (b). Further studies are required to understand this intriguing behavior.


\subsection{Evolution of the gap with U at constant lattice parameters}\label{results_lda_U}


The $5d$ electrons can exhibit different degrees of correlation effects. As explained above, the typical values of the on-site Coulomb repulsion can be on the order of $1-3$ eV. Still, even if the actual U value is difficult to determine accurately, it is interesting to study the evolution of the band gap and the topological properties of the system as a function of the Coulomb on-site repulsion. To take this effect into account, LDA+U calculations were performed for both fillings analysed above retaining a non-magnetic solution.

\subsubsection{LDA+U calculations for the $(\text{Au}_2\text{O}_3)/(\text{Al}_2\text{O}_3)_5$ system}

For the $5d^8$ configuration, the evolution of the gap with the Coulomb interaction was studied for three fixed in-plane lattice parameters: $a_{Al_2O_3} = 4.89$ \AA, $a = 4.76$ \AA\ and $a = 4.95$ \AA. First of all, even though now the calculations include an uncorrelated part treated with LDA, the results are consistent with the previously presented GGA calculations (Fig. \ref{lda_u_au} (a)), the topological phase of those calculations matches the ones obtained now at $\text{U}=0$ using the LDA. The two first lattice parameters (see Fig. \ref{gga_au}) were found to be topological insulators within GGA, and the system with $a = 4.95$ was characterised as a trivial insulator, and so are now at zero U (using LDA).

\begin{figure*}[htc!]
\includegraphics[height=6.5cm]{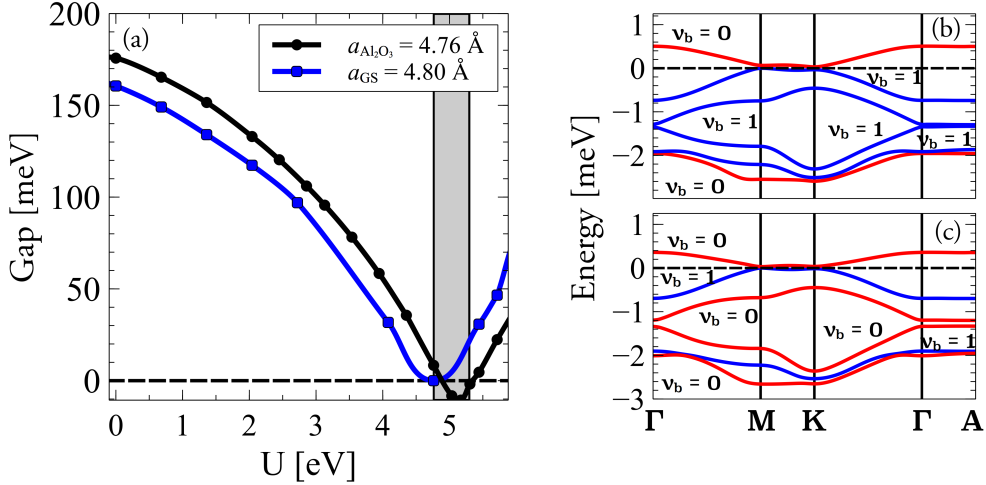}
\caption{(Color online) (a) Evolution of the gap versus the on-site Coulomb repulsion in the $($Os$_2$O$_3)/($Al$_2$O$_3)_5$ system. In contrast with previous studies of $5d^5$ filled honeycomb lattices, no configuration with a non-trivial topology was found with this corundum-based structure. The gap closing and reopening occurs by crossing through an intermediate metallic phase. (b) For the $5d^5$, the evolution of U produces a reordering of the topological invariants of the band structure filling at $a = 4.80$ \AA. All of the LDA+U calculations performed show a non-magnetic solution.}
\label{lda_u_os}
\end{figure*}

The evolution of the gap with the on-site Coulomb repulsion U is shown in Fig. \ref{lda_u_au} (a). The configurations with lattice parameters $a_{Al_2O_3} = 4.89$ \AA\ and $a = 4.76$ \AA\ are topological insulators at U=0 and this topological nontriviality holds until their gaps close and reopen in the interval of $2-3$ eV to a high-U value phase that is a trivial insulator. The configuration of $a = 4.95$ \AA\ is slightly different. At $\text{U} = 0$ it was found to be a trivial insulator, but increasing U the system evolves first into a topological insulator (very near $\text{U} = 0$) behaving in the region of $0.5-3$ eV very similarly to the other two lattice parameters explored. The zero gap point, where the phase transition to a high-U trivial insulator occurs, is shifted to lower U values (from $\text U =2.45$ to $\text U =2.18$ eV), as tensile strain is applied within the range discussed here.

Band structure calculations were performed for $a = 4.76$ \AA\ at both sides of the closing gap and these are shown in Fig. \ref{lda_u_au} (b) (the other two cases are analogous but presenting their phase transitions at different U values). By calculating the $\nu_b$ of each band, we have identified that the system is a topological insulator before the gap becomes zero and as U increases, the gap reopens and the system becomes a trivial insulator at larger U. Introducing static correlations at the LDA+U level does not change drastically the electronic structure. Physically meaningful Coulomb interaction values ($1-3$ eV)  host non-trivial topological phases and larger correlations tend to bring about the disappearance of the topological properties, as a general rule, quite strain-independently.


\subsubsection{LDA+U calculations for the $(\text{Os}_2\text{O}_3)/(\text{Al}_2\text{O}_3)_5$ system}

For the $5d^5$ filling, LDA+U calculations were performed for the lattice parameter of the Al$_{2}$O$_{3}$ substrate $a_{{Al}_2O_3}=4.76$ \AA\ and the ground state $a=4.80$ \AA. Like in the previous GGA calculations and SOC enhacement, no TI state was found. For both cases, the evolution of the gap with U (presented in Fig. \ref{lda_u_os} (a)) is quite similar. The large trivial gap closes and reopens again as a trivial insulating state in the region of $4.7-5.3$ eV. The way the gap does so affects both the \textbf{M} and \textbf{K} points in the band structure, leading to an intermediate metallic phase.

In the ground state configuration, a reordering of the topological invariants of the $t_{2g}$ manifold was found not for bands just around the Fermi level but for the lower lying $t_{2g}$ bands. We can see in Fig. \ref{lda_u_os} (b) the evolution of the ground state on both sides of the transition. A trivial phase is always obtained. However, the topological invariants of each band differ, indicating there is a phase change induced by U, but both phases being trivial. For all physically meaningful U values, the system is in the same trivial insulating phase, the transition occurring at large unphysical correlation strength values.


\section{Summary}\label{conclusion}

All along this work we have shown the conditions that allow the existence of non-trivial insulating phases for a metal-based honeycomb lattice created out of the corundum multilayered $(\text{M}_2\text{O}_3)/(\text{Al}_2\text{O}_3)_5$ system. This is a realization of an oxide-based honeycomb lattice of metal cations that does not require the use of unstable polar surfaces for its growth.

We have studied different $5d$ fillings that could present non-trivial phases by analyzing their band structure around the Fermi level, the existence of Dirac cones and the effects of adding spin-orbit coupling. A Dirac cone near the Fermi level was obtained for the Au-based (at $5d^8$ filling) and also for the Ir-based systems (shifting the chemical potential to one electron less at $5d^5$ filling), which open a gap with SOC.

GGA calculations were performed in order to study the stability of the emerging non-trivial phases with in-plane strain. We observed that the $5d^8$ filling is a TI with the lattice parameters of Al$_2$O$_3$, but with increasing $a$ (tensile strain), the gap closes and reopens in a trivial way. The same procedure was employed in the $5d^5$ filling but no TI phase was found. For this case (Os-based compound) we tested an enhancement of the SOC strength, and yet again all the systems calculated were found to be trivial insulators.

On-site Coulomb repulsion effects for the 5d TM cations in the honeycomb were studied by means of the LDA+U method, for testing the appearance of magnetism and studying the phase diagram in terms of the topological properties of the system. Considering only non-magnetic solutions, the $($Au$_2$O$_3)/($Al$_2$O$_3)_5$ compound presents a phase transition from a low-U TI to a higher-U trivial insulator for values of U between $2-3$ eV. However, the $($Os$_2$O$_3)/($Al$_2$O$_3)_5$ compound was found always in a trivial state, independent of the U value utilized.

We have also studied the robustness of these materials with respect to magnetic ordering. We found that AF solutions are eventually stable for both compounds but they appear at different values of U, our DFT calculations predict that, due to the small band widths of the e$_g$ states, an AF order would be present for every realistic U value $1-3$ eV in the $5d^8$ configuration, whereas in the $5d^5$ system much larger U values (higher than $5$ eV) are necessary to stabilise it. The existence of a magnetic phase based on Au$^{3+}$ cations would be rather unique, caused by the band width reduction due to spatial confinement in those nanostructures.

The results for the time-reversal symmetric phase at $5d^8$ filling in terms of topological properties are similar to those found in buckled honeycomb lattices created out of perovskite bilayers, where non-trivial phases were also predicted. On the other side, the $5d^5$ filling behaves differently, and no topological phases appear. Also the band structures are different, with the Dirac cones appearing at different places in the Brillouin zone. Thus, we can conclude that although in both cases Dirac cones appear at the Fermi level, the different hopping paths in the corundum structure compared to the perovskite bilayers yield a different tight binding model with distinct topological phases.

\acknowledgments

The authors thank the Xunta de Galicia for financial support through project EM 2013/037. V. P. thanks the MINECO of Spain for financial support through the Ramon y Cajal Program and Project No. MAT2013-44673-R. We thank J.L. Lado, D. Baldomir and T. Saha-Dasgupta for fruitful discussions.


\end{document}